# Driven weak to strong pinning crossover in partially nanopatterned 2H-NbSe$_2$ single crystal

**Gorky Shaw, Jaivardhan Sinha, Shyam Mohan and S. S. Banerjee**
Department of Physics, Indian Institute of Technology, Kanpur-208016, U. P., India

E-mail: satyajit@iitk.ac.in

**Abstract:** Investigations into the heterogeneous pinning properties of the vortex state created by partially nano-patterning single crystals of 2H-NbSe$_2$ reveal an atypical magnetization response which is significantly drive dependent. Analysis of the magnetization response shows non-monotonic behavior of the magnetization relaxation rate with varying magnetic field sweep rate. With all the patterned pinning centers saturated with vortices, we find that the pinning force experienced by the vortices continues to increase with increasing drive. Our studies reveal an unconventional dynamic weak to strong pinning crossover where the flow of the vortex state appears to be hindered or jammed as it is driven harder through the interstitial voids in the patterned pinning lattice.

## 1. Introduction

The vortex state in superconductors is known to exhibit a variety of glassy phases like the Bragg glass, the vortex glass phase in the presence of random point pins, and the Bose glass phase in the presence of extended columnar defects [1]. These vortex phases are associated with a homogenous pinning environment. An important and technologically relevant area of investigation is the nature of vortex state in presence of pinning centers generated by nanoscale patterning of a superconductor [2,3,4,5,6,7,8,9,10,11,12]. Spatial dimension of these patterned pinning centers can range from sizes comparable to the superconducting coherence length ($\xi$) to the penetration depth ($\lambda$). Unlike most previous studies which have been on superconductors with homogeneous pinning [1], here, we investigate the case of a sample with heterogeneous pinning generated with nanopatterning. Heterogeneous pinning consists of two distinct pinning populations, viz., one consisting of randomly distributed weak pins and another of spatially correlated nanopatterned strong pins. From the magnetization response measurement of our partially patterned superconductor we find that contrary to expectation, at low drives rather than the vortex state being pinned strongly on the blind hole lattice it favors a weakly pinning state. However under the influence of a larger drive the vortex state transforms from a weak to strong pinning state. This crossover from weak to strong pinning doesn't occur as a natural consequence of changes in the equilibrium properties of the vortex state, like that associated with softening of the elastic vortex state [1,13,14]. Instead we find that the crossover is achieved only by driving the vortex state. To understand the drive dependent weak to strong pinning crossover we propose that the vortices experience jamming as they are driven hard through the blind hole lattice.

## 2. Experimental details, results and discussion

### 2.1. Sample Preparation

A number of recent studies on nanopatterned superconductors have been on thin films [10,11], unlike this letter which reports on studies of nanopatterned high quality, weak pinning single crystals. In comparison to thin films, single crystals of a low-$T_c$ superconductor, 2H-NbSe$_2$ possess very low [15] $J_c/J_0 \sim 10^{-6}$ (where $J_c$ is the critical current density of the superconductor which is proportional to the pinning strength and $J_0$ is the depairing current density). A low $J_c/J_0$, implies weak pinning environment in the NbSe$_2$ single crystals which affords an easy comparison of changes in pinning produced by nanopatterning unlike that in thin films. We have investigated two NbSe$_2$ crystals belonging to a batch [14] of high quality single crystals (with dimensions ~ 1.5 *mm* x 1 *mm* x 30 $\mu m$, and $T_c(0) = 7\ K$). The two crystals are obtained from the same NbSe$_2$ single crystal by cleaving it with scotch tape. One of the NbSe$_2$ crystals was milled with a focused Ga ion beam (diameter ~ 7 *nm*) using the Focused Ion Beam (FIB) machine (dual beam FEI make Nova 600 NanoLab) to produce a hexagonal array of blind holes. Each hole has a diameter of 170 *nm* ( > $2\xi_{ab}$ ~ 15.4 *nm* and ≈ $\lambda_{ab}$ ~ 120 *nm* for the 'ab' crystal orientation in NbSe$_2$ [16]) and a mean center-to-center spacing between the holes (d) of 350 *nm* and depth of ~ 1$\mu m$ covering a rectangular area of 180 $\mu m^2$ (with about 1660 holes, in the patterned region which is located away from the sample edges). Inset of figure 1 shows a Scanning Electron Microscopic (SEM) image of the magnified portion of the patterned area.

### 2.2. Magnetization measurements

Shown in Fig.1 main panel are the bulk dc magnetization hysteresis *M(H)* loops for the patterned and unpatterned NbSe$_2$ samples measured at T = 5 *K* with a Quantum Design Superconducting Quantum Interference Device (SQUID) magnetometer. To avoid effects due to dissimilar sample geometry, the two samples measured were matched in physical dimensions and shape. Figure 1 main panel shows that the *M(H)* for the patterned and unpatterned samples are nearly identical at H > 0.02 *T*. As the width of the magnetization hysteresis ($\Delta M$) is a direct measure of the pinning strength in the superconductor [17], it appears that due to nanopatterning with blind holes, the pinning of vortices is only weakly affected.

At this juncture, it is worthwhile recalling a few important length and magnetic field scales defined by nanopatterning are: Spacing (*d*) between the blind holes of 350 *nm*, implies that at $B_\Phi \sim \dfrac{\phi_o}{d^2} = 0.0195\ T$ (where $\Phi_o = 2.07\times 10^{-15}\ Wb$ and $B_\Phi$ is the matching field value) there exists on average one vortex per blind hole. In this paper the field regime of our investigation in NbSe$_2$ is at magnetic fields (B) > 0.0195 *T*. Apart from $B_\Phi$, another relevant field range for our studies is the saturation field $H_s$. We estimate that in NbSe$_2$ within each blind hole of radius $r = 85$ nm, $n_s = r/2\xi_{ab}$ ~ 5 to 6 vortices can be accommodated. Using $n_s$ ~ 5 and 1660 blind holes patterned in an area of 180 $\mu m^2$ in NbSe$_2$, at $H_S = \frac{n_s \times 1660}{180 \mu m^2} \times \phi_0$ ~ 0.09 – 0.1 $T$ all blind holes are nearly saturated with vortices (cf. arrow marked H$_s$ in Fig.2). We expect that at $H > H_s$, if most of the blind holes are saturated with vortices then we should be able to identify a distinct response of vortices pinned on strong blind holes pinning sites, from those which are not. Infact, the magnetization relaxation behavior of the vortices pinned on the blind holes is expected to be distinct from those pinned on weak pins. Therefore to effectively distinguish between different pinned phases, we investigate the behavior of *M(H)* loops for different magnetic field (*H*) sweep rates β ~ 0.005, 0.05, 0.1, 0.2 and 0.3 *T/min* using the Vibrating Sample Magnetometer (VSM) (Oxford make, Model 3001). In Fig.2 main panel, we compare the *M(H)* loops at T = 5 *K* for the patterned sample recorded with β of 0.005 *T/min* and 0.2 *T/min*. Inset (a) of Fig.2 shows that at 5 K at H = 0.2 *T* (> B$_\Phi$, H$_s$), the behavior of $\Delta M_{patt}/\Delta M_{unpatt}$ versus β, where $\Delta M_{patt}$ and $\Delta M_{unpatt}$ are the widths of the *M(H)* loops for the patterned and unpatterned sample recorded at the same β. We observe that for β < β′ (where β′ is marked in the Fig.2 inset (a)), the $\Delta M_{patt}/\Delta M_{unpatt} \approx 1$, implying that the hysteresis curve behave almost identically for low β. Beyond the threshold value of β′ ~ 0.05 *T/min*, the ratio $\Delta M_{patt}/\Delta M_{unpatt}$ increases significantly by a factor of almost 5. This indicates that beyond a threshold β′ ~ 0.05 *T/min* the increase in the width of M(H) loop for the patterned sample is much more in comparison to the increase in the unpatterned sample (at the same β). As the width ΔM is indicative of pinning, it is interesting to note that in terms of pinning, from the above data it appears that only beyond a threshold β′ ~ 0.05 *T/min*, vortices are driven into a state with strong pinning (larger ΔM(β)).

The above atypical sweep rate dependent magnetization response in the patterned sample is also analyzed in terms of relaxations due to thermally activated creep. Usually the value of the magnetic moment depends on the strength of the shielding currents and hence on the sweep rate of the magnetic field (equivalent to an effective electric field). In such a situation one can define the magnetization relaxation rate with $R = \pm dM/d(Log\beta)$ [18], where '+' and '-' correspond to the descending and ascending branches of M(H) loop. Inset (b) of figure 2 shows at H = 0.2 *T* at 5 *K*, a compilation of the R(β) from the *M(H)* curves for the patterned sample. In this inset we see that in the patterned sample the value of R remains nominally positive at +0.05 until the threshold β′ ~ 0.05 *T/min*. In Fig.2(b) we observe a significant upturn in R, viz., R increases from 0.05 to +0.5 for β > β′ which also coincides with the β where *ΔM$_{patt}$(β)* increases w.r.t. *ΔM$_{unpatt}$(β)* as seen in fig.2(a). We also observe an upward shift of the irreversibility field *H$_{irr}$(T)* for the patterned sample at β = 0.2 *T/min* (one representative value of *H$_{irr}$(T)* shown in figure 2 main panel) confirming that a larger pinning landscape is probed at large β beyond a threshold value of β′ ~ 0.05 *T/min*. As argued in Ref.18, the relaxation rate $R \propto \frac{1}{\rho_d}$, where $\rho_d$ is the differential resistivity of the vortex state.

An enhancement in R (beyond β′) reflects a significant decrease in $\rho_d$ precipitated by an abrupt enhancement in pinning experienced by the vortex state in the patterned sample when the field sweep rate exceeds a threshold viz., β′. It is important to note that the increase in pinning experienced by the vortex state at large β (leading to reduced $\rho_d$) is found at magnetic fields of above 0.2 $T$ which is well above $H_s \sim O(0.1\ T)$, viz., at magnetic fields where the density of vortices is such that all the strong pinning blind holes have already been saturated with vortices. Earlier studies on unpatterned 2H-NbSe$_2$ single crystals have shown that in a certain field temperature regime close to the upper critical field, due to the peak effect (PE) phenomenon[14,15] there is a significant enhancement in pinning. The PE phenomenon is associated with softening of the vortex state well before pinning vanishes in the sample at $H_{irr}$ leading to a strongly pinned vortex state. Based on studies on our present batch of crystals [14,15] we know that our present field – temperature regime where we observe a drive dependent weak to strong pinning crossover is not coincident with the PE phenomenon regime. Infact later on in Fig.5 inset we show that a measurement of the pinning force in the field temperature regime of our interest in the unpatterned 2H-NbSe$_2$ sample does not show any evidence of PE. We surmise that the drive dependent crossover in pinning in our partially patterned sample has an origin distinct from effects related to softening of the vortex state.

*2.3. Geometric effects and the magnetization response*

In superconductors with weak bulk pinning, like in 2H-NbSe$_2$, it is well known that the surfaces of superconductors play a significant role in inhibiting the entry and exit of vortices from the sample edges. These edge effects are known to be significant in samples with weak bulk pinning leading to significant modification in the shape and behavior of irreversible magnetization response of the superconductors[19,20] (which are not generated due to bulk pinning). A common signature associated with surface barriers is the reverse leg of magnetization in the M-H hysteresis loop remains nearly constant over a wide field range (see N. Chikumoto et al in Ref.19 and P. K. Mishra et al in Ref.20). Our figures 1 and 2 exhibit significant field dependence of the reverse magnetization in both our samples, indicating that geometric barriers effects are insignificant in our patterned and unpatterned 2H-NbSe$_2$ samples. It was shown [19,21] that in the presence of barrier effects, at fields greater than the penetration field, the width of the magnetization hysteresis loop (using a constant $M$ in the reverse leg) varies as, $\Delta M \propto \frac{1}{H}$ (such behavior has been observed for 2H-NbSe$_2$ in [20]). In the inset of Fig. 3(b) we have plotted ΔM versus $\frac{1}{H}$ obtained from the M-H data for both the patterned and unpatterned samples at 5 $K$ and 6 $K$. It is clear from the nonlinear curve in inset Fig.3(b) that geometric effects are insignificant.

In the main panels (a) and (b) of Fig.3, we show the in-phase ($\chi^{/}$) and out of phase component ($\chi^{//}$) of the ac – susceptibility response measured as a function of T in a dc field of $H_{dc}$ = 0.1 $T$ with a superimposed probing ac field ($h_{ac}$) of $2 \times 10^{-4}$ $T$ (frequency 211 Hz) on the 2H-NbSe$_2$ crystals prior to patterning. From Fig.3(a) we see that at 5 $K$ (which is the same temperature at which our results have been reported in this paper), the diamagnetic shielding response (viz., $\chi^{/}$) is close to -0.6 which is well above -1 (associated with a perfect diamagnetic screening response). If geometric barriers were to dominate the shielding response at 5 $K$ and 0.1 $T$, then vortices would have been strongly shielded out from the interior of the sample, and we would have found the diamagnetic shielding response viz., $\chi'$ approaching a value close to -1. Such a behavior is absent in our samples. Another significant observation is, Fig.3(b) shows that the dissipation response (viz., $\chi''$) at 5 *K* and 0.1 *T* is far from zero. The significantly large dissipation response (which is well above zero) at 5 *K* and 0.1 *T* indicates that the dissipation emanates from the bulk of the sample due to the full penetration at $h_{ac}$ field into the superconductor [22]. If geometric barriers were to be significant at these fields and temperatures, then along with a large $\chi'$, due to insufficient penetration $\chi''$ should also have achieved a value close to zero, which is absent in Fig.3(b) at 5 *K*. Thus not only do we find absence of distinct signatures of geometric effects but these effects are also expected to become weak at large fields. Hence in the field range in which we discuss our results (i.e., at $H > H_s \sim 0.1$ T), geometric barriers have an insignificant role to play in determining the behavior of magnetization response of the sample.

*2.4. E-J calculation*

While our SQUID magnetization measurements reveal information about the static vortex state through the irreversible magnetization response, the fast sweep rate dependent M-H response gives the integrated magnetization response emanating from a driven vortex state. As only a small area of the sample has been patterned (0.02%), therefore the static magnetization response associated with the pinned vortex state in the patterned sample is only weakly affected. We confirm this from Fig.1 where the difference in the magnetization response between the patterned and unpatterned sample is small. However the discussions associated with Fig.2 show that the integrated magnetization response of the driven vortex state which is flowing in between the saturated ($H > H_s$) blind hole pins, is significantly different from a driven vortex state response in a sample without the blind holes. The sweeping of H, drives the weakly pinned vortices (in the unpatterned regions of the sample and those in the interstitial spaces between the blind holes) to reorganize themselves inside the superconductor when the induced screening current (J) exceeds the $J_c$. The driven vortices generate an electric field of magnitude $E = uB$, where $u$ is the velocity of the vortices. An approximate procedure used to construct the electric field from the M(β) data [23,24] is via $E \propto A \times (dM/dH)/(dH/dt)$; (where *A* is the sample area). The width of the hysteresis loop ($\Delta M$) provides information about the shielding current *J* induced in the superconductor as *H* is swept ($(\Delta M / w) \propto J$, $w$: mean width of the sample [17]). To analyze the *E - J* curves we use a relationship invoked in the context of collective vortex creep in vortex glasses [24,25] viz.,

$E \propto \exp\left[-\left(U_c/kT\right)\left(J_c/J\right)^\mu\right]$; where $U_c$ is the depth of the potential well at $J = J_c$. Figure 4 compares the *E - J* curve determined from forward leg of *M(H)* curve recorded with β = 0.05, 0.1 and 0.2 *T/min* for the patterned and unpatterned samples. The dashed (red) and solid (blue) lines correspond to the *E(J)* fitted to the data using the above expression with a fixed $\mu$ = 3 and $\mu$ = 1 respectively, with $U_c$ and $J_c$ as variable fitting parameters. For the unpatterned sample the E-J curve fits to $\mu = 1$ over the entire E-J range even at 0.2 *T/min*. From Fig.2 main panel we know that J $\propto$ $\Delta M$ decreases monotonically with increasing H for the patterned sample at 0.2 *T/min*. Thus, the observed change in the shape of the E-J curve in fig.4 for the patterned sample (viz., a fit with $\mu = 1$ changes to a fit with $\mu = 3$) cannot be attributed to any non-monotonic variation in the J(H) intrinsic to 2H-NbSe$_2$ (for example the behavior in J$_c$(H) found in the vicinity of the PE phenomenon [15]). The fitted solid (blue) line extrapolates to $J_{c,in}$ while the fitted dashed (red) curve extrapolates to $J_{c,bh}$, where E = 0. In general, note that for the patterned sample the *E(J)* for β ≤ a range of 0.05-0.1 *T/min* (~ β$'$) has the lower $J_{c,in}$ intercept. Theory predicts that $\mu \sim 1$ to 1.5 for weak collective pinning (see M. V. Feigel'man et al. Ref. 25). Thus, the source of the lower intercept, $J_{c,in} < 1\times10^3$ $A/m^2$ is the weakly pinned vortex phase with $\mu = 1$. The value of $\mu = 3$ at $\beta > \beta'$ indicates a departure from weak collective creep scenario. At $\mu = 3$ the $J_{c,bh} \geq 2\times10^3$ $A/m^2$ intercept is higher (compared to that at $\mu = 1$) implying the E-J behavior corresponds to the onset of a strongly pinned phase. Thus at $\beta < \beta'$ the response observed in the patterned sample emanates from weak collective pining which transforms to a strong pinning regime at $\beta > \beta'$ where the E-J curve fits with $\mu = 3$. In general, neutron diffraction studies have shown that any change in curvature of the I-V and equivalently the E-J characteristics associated with the driven vortex state in a superconductor is associated with a change in the degree of inhomogenity in the distribution of currents flowing in the sample [26]. Thus, changes we observe in the E – J curvature (viz., a change from $\mu = 1$ to $\mu = 3$ fit) indicates that perhaps the nature of the current distribution in the patterned sample undergoes significant changes as $\beta$ changes from $< \beta'$ to $> \beta'$ concomitant with a crossover in pinning a weak (with low $J_{c,in}$) to a strong (with a higher $J_{c,bh}$) pinning state at large drives.

*2.5. Determination of irreversibility line and pinning force*
In the inset of Fig.5, we compute the pinning force curve at T = 5 K (using, $F_p(H) \propto \Delta M \times H$; as $J \propto \Delta M$) for the unpatterned sample (recorded at *0.3 T/min*, black dashed curve) and the patterned sample (at different sweep rates, solid (colored) curves). It is clear that the $F_p(H)$ curves for the patterned sample at low sweep rates ($\leq 0.1$ *T/min*, viz., for $\beta \leq \beta'$) coincides with that for the unpatterned sample (recorded at 0.3 *T/min*, black dashed curve). This indicates that the nature of pinning is weak at low $\beta$ in the patterned sample. At low $\beta$, the $F_p(H)$ displays a peak like feature near 0.2 *T* (far away from $H_{c2}$). Such a peak like feature in the pinning force curve is well known [27] and is associated with weak two-dimensional collective pinning in 2H-NbSe$_2$. However at larger $\beta$ (0.2 and 0.3 *T/min*) one observes that in the vicinity of 0.2 *T* and beyond, the $F_p(H)$ curves for the partially patterned sample begins to deviating significantly from that for the unpatterned sample. In fact for $H > 0.4$ *T* ($> H_s = 0.1$ *T*) for the patterned sample, at a $\beta$ of 0.3 *T/min*, the pinning force is few orders of magnitude larger than the pinning from unpatterned sample at the same field. At low $\beta$ ($\leq \beta'$) below 0.2 *T*, the coincidence of the $F_p(H)$ curves for the patterned and unpatterned sample indicates that the pinning in the patterned sample is being governed by the naturally occurring weak pinning centers present in 2H-NbSe$_2$. At large $\beta$ (0.2 and 0.3 *T/min*) beyond 0.2 *T*, we see a clear departure from the weak collective pinning behavior present for low $\beta$ (below 0.1 *T/min*). In the inset of Fig.5 we show a dashed vertical line demarcating the crossover from weak collective pinning to strong pinning beyond 0.2 *T* at large $\beta$ ($>\beta'$) in our partially nano-patterned sample of 2H-NbSe$_2$.

To understand what results in a stronger pining at large $\beta$, we investigate the behavior of the irreversibility line ($H_{irr}(T)$). It is clear from the fig.2 main panel that in the patterned sample at 5 K at 0.2 *T/min* (the open circles) the M-H loop has a higher $H_{irr}(T)$ as compared to the same loop recorded with $\beta = 0.005$ *T/min* ($H_{irr}(T)$) is that field at which the M(H) becomes path independent viz., $M_{for}$ and $M_{rev}$ coincide). The main panel of Fig.5, shows the comparison of $H_{irr}(T)$ line, for the patterned sample measured from hysteresis loops recorded with the field being swept at 0.2 *T/min* (solid (red) line) ($H_{irr,0.2T/min}$) and for hysteresis loops measured at 0.005 *T/min* (dotted (blue) line) (0.005 *T/min* has been abbreviated to 0T/min and denoted as $H_{irr,0T/min}$ for convenience). Note that Fig.5 shows, $H_{irr,0T/min}$ for the patterned sample is identical to that for the unpatterned one(cf. solid and open (blue) squares respectively), confirming the weak collective pinning behavior dominates at low β. At high β, the crossover to the strong pinning regime coincides with $H_{irr,0.2T/min}$ to be shifted upwards w.r.t $H_{irr,0T/min}$. The observation that the $H_{irr}(T)$ is shifted up w.r.t to the line in the unpatterned sample is reminiscent of a similar trend (relating, only to the upward shift trend and not to the β dependence) of the irreversibility line found in samples with extended defects like columnar defects produced with heavy ion irradiation [28].

*2.6. Discussion*
At H which is greater than but close to $H_s$ (the saturation field, where on average all blind holes are occupied by vortices), vortices strongly pinned on periodically spaced blind hole pinning centers creates a rigid cage like potential due to the regular hexagonal arrangement of the blind hole lattice. This cage confines the vortices present in the interstitials between the blind holes at $H > H_s$. Due to this trapping inside the cage, the thermally induced wandering of the vortex line from their mean position is suppressed thereby leading to an upward shift of $H_{irr}(T)$. The upward shift of $H_{irr}(T)$ explained via the cage model, had been proposed in the past to explain features for columnar defects [29]. While the caging potential created by the vortices strongly pinned in the blind holes explains the upward shift of $H_{irr}(T)$, it does not provide an explanation of the sweep rate dependence of the pinning force which increases with β especially for $H \gg H_s$. From inset of Fig.5 it appears that at the same H ($> H_s$), viz., for example at say 0.8 T (~ $40B_\phi$ and $8H_s$), for the same density of vortices the pinning force depends on the sweep rate at which the field is reached. At $H \gg H_s$, the density of vortices is such that most of the blind holes are already saturated with strongly pinned vortices and the excess vortices in the interstitials of the blind hole lattice are caged by the strongly pinned blind holes. However at large vortex density the repulsion between vortices is very high and it is unlikely that the cage confinement scenario will be significantly effective upto $H \gg H_s$. Given the above situation, one expects the pinning force experienced by the vortex state to depend on the density of vortices, rather than dependent of the rate at which one achieves the given density of vortices. In $NbSe_2$ at $H > H_s$, the regime where R(β) response becomes significantly +ve viz., at β >β′ (cf. inset fig.2(b)), it appears that despite a strong caging potential the pinning strength enhances with β (beyond a threshold β′).

We propose that the vortices present in the voids of the blind hole lattice cannot be driven through the interstitial spaces in the lattice uniformly at large drives. Rather than being uniformly driven faster at larger drives a non monotonic behavior sets in, wherein the vortices begin to become more immobile at larger drives. This effect becomes more pronounced with increasing β. At this juncture drawing upon a scenario based on a recent simulation [30] we propose that our observation of an enhancement in pinning strength is a result of a jamming phenomenon when vortices are driven across the interstitial voids in the blind hole lattice at large sweep rates. It has been shown [31] that, at 1.5 $B_\phi$ < B, cluster states of n-vortices ('n-mer states') form in the interstitials voids between the patterned pinning sites. These n-mer's crystallize into ordered structures also termed as vortex molecular crystals. It is found that [30] under the influence of a drive applied in a particular direction w.r.t to the pinning lattice, the n-mers flow is misaligned to the drive in certain directions while along other directions they are aligned. This type of a flow leads to a jamming of the driven vortex molecular crystalline state. For a driven elastic vortex state, if any region in the vortex medium experiences strong pinning, then the entire elastic vortex medium (due to elasticity) collectively slows down. We believe this happens in our patterned sample. Due to local jamming effects encountered in the interstitials located in between the blind holes in the partially patterned region of the sample there is a slowing down of the entire weakly pinned driven elastic vortex medium. Such a slowing down of the vortex state results in the drive dependence of the bulk magnetization response we observe in our partially patterned sample. This driven jammed state is characterized by a strong enhancement in the pinning force. In our study as the applied driving force on the vortex state is directionally symmetric w.r.t to the blind hole pinning lattice, the probability of having jamming caused by aligned and misaligned n-mer flowing vortex states is high. Higher the $\beta$, greater is the immobilization of the vortex state due to jamming. We propose that the observed enhancement in pinning force at large $\beta$ is a consequence of jamming of n-mer vortex state driven through the interstitial voids in between the blind hole lattice. At larger driving force (viz., larger $\beta$) jamming causes an enhancement in the pinning force. The observation of a larger $J_{c,bh}$ at higher $\beta$ in Fig.4, is consistent with the enhancement in pinning force due to jamming. From Fig.4, at $J > J_{c,bh}$ the dE/dJ, which is proportional to the differential resistivity of the driven vortex state, shows that initially (for J in the vicinity of $J_{c,bh}$), it increases and at higher J becomes smaller and approaches a saturated value. Studies on plastically driven vortex state, where there are mobile channels of freely flowing vortices with islands of strongly pinned vortices exhibit a similar feature in differential resistance [15,32], viz., a convex curvature of I-V, where the differential resistance (dV/dI) value exhibits an initial peak which returns into a saturated value. It is interesting to note, that unlike what is found in unpatterned samples of 2H-NbSe$_2$ where the weak to strong pinning crossovers occur due to transformations in the equilibrium properties of the vortex state, like, vortex state softening [1,13, 14,15], here the crossover is a dynamic one from a weak to strong pinning regime. As a consequence of jamming of vortices in the interstitials between the blind hole lattice, there will be a significant barrier for the entry of extra vortices into this patterned area. A preliminary evidence for this barrier is presented elsewhere [33]. More detailed experiments are underway to investigate the weak to strong pinning crossover at large drives in the patterned blind hole lattice.

## 3. Conclusion

In conclusion, by partially patterning the superconductor we have created a heterogeneously pinned vortex phase, consisting of a weakly pinned fraction of vortices in the unpatterned regions and another fraction of strongly pinned vortices in the nanopatterned region. The unique feature of the partially patterned superconductor is that the weak and strong pinning phases do not occur as a natural outcome of peculiarities of the vortex state, like due to softening of the vortex state. We propose that our observation of a dynamic weak to strong pinning state crossover at field much larger than the saturation field of the blind holes, is due to a jamming of the vortex state which is driven through the interstitial voids present in-between the blind hole lattice. Our future studies are underway to investigate the static and dynamic phase of the interstitial vortex states present in-between the blind hole lattice.

**Acknowledgments**

SSB would like to acknowledge the funding support from DST- India, IIT Kanpur, India and CSIR-India. SSB thanks Professor P L Paulose and Professor A K Grover of TIFR, Mumbai and Professor Eli Zeldov of Weizmann Institute of Science, Israel, for fruitful discussions. We gratefully acknowledge the experimental support from Mr. Pabitra Mandal.

[1] Blatter G, Fiegel'man M V, Geshkenbein V B, Larkin A I and Vinokur V M 1994 *Rev. Mod. Phys.* **66** 1125
Giamarchi T and Le Doussal P 1996 Phys. Rev. Lett. **76** 3408
[2] Kanda A, Baelus B J, Peeters F M, Kadowaki K and Ootuka Y 2004 *Phys. Rev. Lett.* **93** 257002
Geim A K, Dubonos S V, Lok J G S, Henini M and Maan J C 2000 *Nature* **396** 144
Palacios J J 2000 *Phys. Rev. Lett.* **84** 1796
[3] Fiory A T, Hebard A F andSomekh A F 1978 *Appl. Phys. Lett.* **32** 73
[4] Baert M, Metlushko V V, Jonckheere R, Moshchalkov V V and Bruynseraede Y 1995 *Phys. Rev. Lett.* **74**, 3269 (1995), *ibid*, *Europhys. Lett.* **29,**157 (1995)
[5] Silhanek A V, Raedts S, Van Bael M J and Moshchalkov V V 2004 *Phys. Rev. B*. **70** 054515
[6] Zhukov A A, de Groot P A J, Metlushko V V and Ilic B 2003 *Appl. Phys. Lett.* **83** 4217
[7] Geim A K, Dubonos S V, Palacios J J, Grigorieva I V, Henini M and Schermer J J 2000 *Phys. Rev. Lett* **85** 1528
[8] Chibotaru L F, Ceulemans Arnout, Bruyndoncx Vital and Moshchalkov Victor V 2000 *Nature* **408** 833
[9] Karapetrov G, Fedor J, Iavarone M, Rosenmann D and Kwok W K 2005 *Phys. Rev. Lett.* **95** 167002
[10] Raedts S, Silhanek A V, Van Bael M J and Moshchalkov V V 2004 *Phys. Rev. B* **70** 024509
[11] Moshchalkov V V, Baert M, Rosseel E, Metlushko V V, Van Bael M J and Bruynseraede Y 1997 *Physica C* **282** 379
[12]Goldberg S, Segev Y, Myasoedov Y, Gutman I, Avraham N, Rappaport M, Zeldov E, Tamegai T, Hicks C W and Moler K A 2009 Phys. Rev. B **79** 064523
[13] Blatter G, Geshkenbein V B and Koopmann J A G 2004 *Phys. Rev. Lett.* **92** 067009
[14] Mohan Shyam, Sinha Jaivardhan, Banerjee S S and Myasoedov Yuri 2007 *Phys. Rev. Lett.* **98** 027003
[15] Higgins M. J. and Bhattacharya S 1996 *Physica C* **25** 232
Kes P H 1995 *Nature* **376** 729.
[16] 2000 *Handbook of Superconductivity* ed Charles P. Poole, Jr. (Academic press, Elsevier, London, UK)
Sonier J E, Kiefl R F, Brewer J H, Chakhalian J, Dunsiger S R, MacFarlane W A, Miller R I, Wong A, Luke G M and Brill J W 1997 *Phys. Rev. Lett.* **79** 1742
[17] Bean C P 1964 *Rev. Mod. Phys.* **36** 31
[18] Zhukov A A, Kokkaliaris S, de Groot P A J, Higgins J, Bhattacharya S, Gagnon R and Taillefer L 2000 Phys. Rev. B **61** R886
[19] Bean C P and Livingston J D 1964 *Phys. Rev. Lett.* **12** 14
Campbell A M and Evetts J E 1972 Critical Currents in Superconductors (Taylor & Francis, London) p. 140
Clem J R 1974 in: *Proc. of Low Temperature Physics* LT 13, vol. 3, eds. Timmerhans K D *et al.* (Plenum, New York) p. 102.
Konczykowski M, Burlachkov L I, Yeshurun Y and Holtzberg E 1991 *Phys. Rev. B* **43** 13707
Chikumoto N, Konczykowski M, Motohira N and Malozemoff A P 1992 *Phys. Rev. Lett.* **69** 1260.

Zeldov E, Larkin A I, Geshkenbein V B, Konczykowski M, Majer D, Khaykovich B, Vinokur V M and Shtrikman H 1994 *Phys. Rev. Lett.* **73** 1428
[20] Mishra P K, Ravikumar G, Sahni V C, Koblischka M R and Grover A K 1996 *Physica C* **269** 71.
Mishra P K, Ravikumar G, Chandrasekhar Rao T V, Sahni V C, Banerjee S S, Ramakrishnan S, Grover A K and Higgins M J 2000 *Physica C* **340** 65
[21] Burlachkov L, Geshkenbein V B, Koshlev A E, Larkin A I and Vinokur V M, 1994 *Phys. Rev. B* **50**, 16770.
[22] Clem J R in Magnetic Susceptibility of Superconductors and other Spin Systems, edited by R. A. Hein, T. L. Francavilla and D. H,. Leibenberg (Plenum press, New York, 1991), p.177
Ling X. S. and Budnick J., *ibid*, p377
Angurel L A, Amin F, Polichetti M, Aarts J and Kes P H, 1997 *Phys. Rev. B* **56** 3425.
[23] Kupfer H, Gordeev S N, Jahn W, Kresse R, Meier-Hirmer R, Wolf T, Zhukov A A, Salama K and Lee D 1994 *Phys. Rev. B* **50** 7016
Maley M P, Kung P J, Coulter J Y, Carter W L, Riley G N and McHenry M E 1992 *Phys. Rev. B* **45** 7566
[24] Sandvold E and Rossel C 1992 *Physica C* **190** 309
Konczykowski M, Vinokur V M, Rullier-Albenque F, Yeshurun Y, Holtzberg F 1993 *Phys. Rev. B* **47** 5531
[25] Fisher Matthew P A 1989 *Phys. Rev. Lett.* **62** 1415
Fisher Daniel S, Fisher Matthew P A and Huse David A 1991 *Phys. Rev. B* **43** 130
Feigel'man M V, Geshkenbein V B and Vinokur V M 1991 *Phys. Rev. B* **43** 6263
Feigel'man M V, Geshkenbein V B, Larkin A I and Vinokur V M 1989 *Phys. Rev. Lett* **63** 2303
[26] Pautrat A, Goupil C, Simon Ch, Charalambous D, Forgan E M, Lazard G and Mathieu P 2003 *Phys. Rev. Lett.* **90** 087002
Pautrat A, Scola J, Simon Ch, Mathieu P, Brûlet A, Goupil C, Higgins M J and Bhattacharya S 2005 *Phys. Rev. B* **71** 064517
[27] Koorevaar P, Aarts A, Berghuis P and Kes P H 1990 *Phys. Rev. B* **42** 1004
[28] Civale L, Marwick A D, Worthington T K, Kirk M A, Thompson J R, Krusin-Elbaum L, Sun Y, Clem J R and Holtzberg F 1991 *Phys. Rev. Lett.* **67** 648
Banerjee S S, Soibel A, Myasoedov Y, Rappaport M, Zeldov E, Menghini M, Fasano Y, de la Cruz F, van der Beek C J, Konczykowski M and Tamegai T 2003 *Phys. Rev. Lett.* **90** 087004
[29] Nelson D R and Vinokur V M 1993 *Phys. Rev. B* **48** 13060
Vinokur V, Khaykovich B, Zeldov E, Konczykowski M, Doyle R A and Kes P H 1998 *Physica C* **295** 209
[30] Reichhardt C and Olson Reichhardt C J 2008 *Phys. Rev. B* **78** 224511
[31] Reichhardt C and Olson Reichhardt C J 2007 *Phys. Rev. B* **76** 064523.
[32] Olive E and Soret J C 2006 *Phys. Rev. Lett.* **96** 027002
*ibid* 2008 *Phys. Rev. B* **77** 144514
[33] Banerjee S S, Shaw Gorky, Sinha Jaivardhan, Mohan Shyam and Mandal Pabitra 2009 *Physica C* (in press; doi:10.1016/j.physc.2009.11.096)

**Figures:**

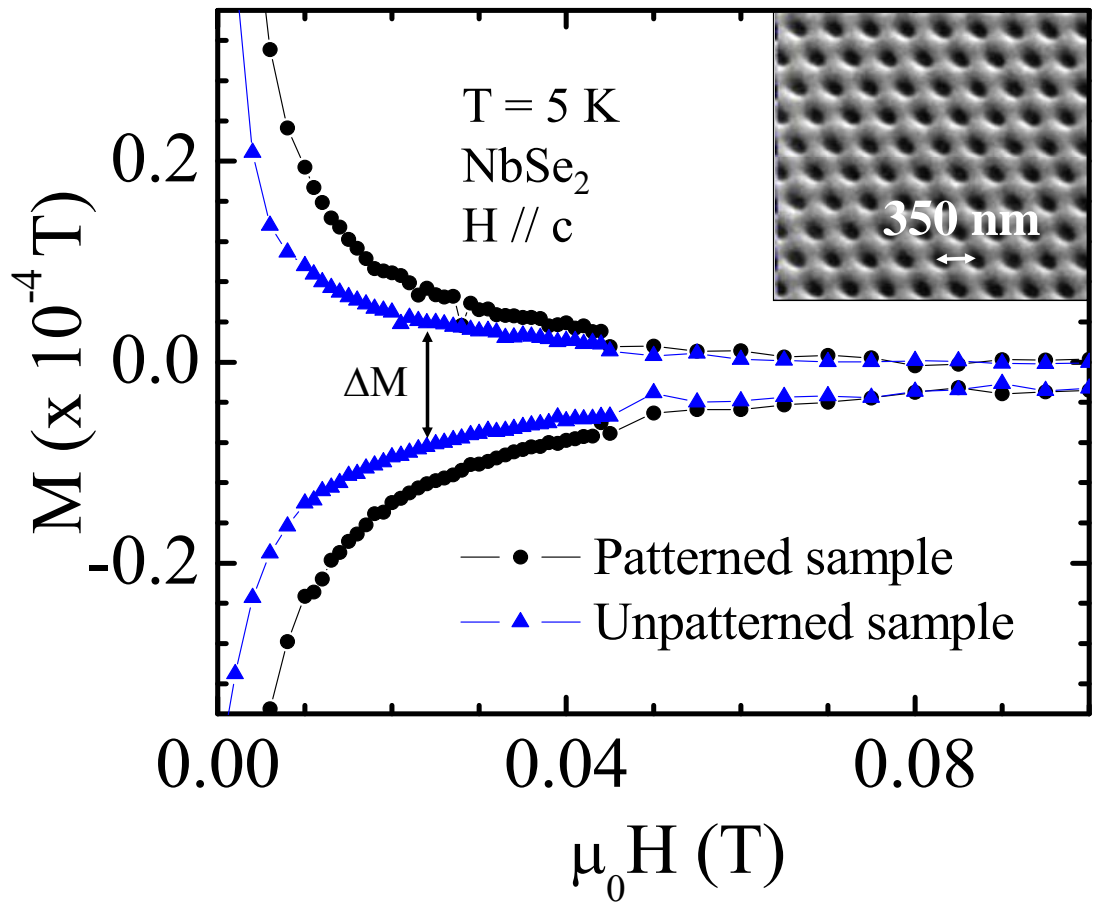

**Fig.1**: Isothermal bulk dc magnetization hysteresis *M(H)* loops of the patterned and unpatterned NbSe$_2$ samples (denoted by circles and triangles respectively) measured at 5 *K*. Inset shows an SEM image of the region of the NbSe$_2$ crystal patterned with a hexagonal array of blind holes.

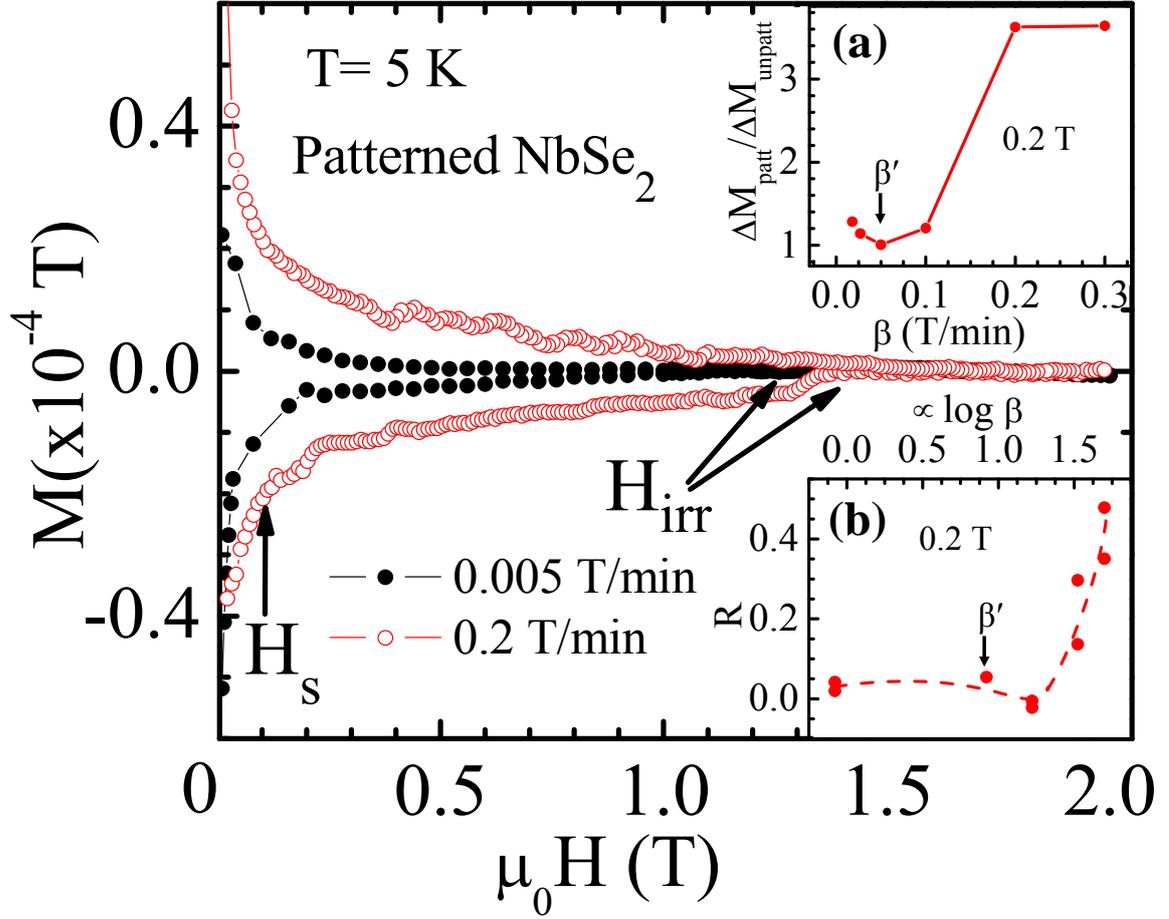

**Fig.2:** The main panel compares at 5 *K* the *M(H)* loops for the patterned sample recorded with β = 0.005 *T/min* (black curve, closed circles) and β = 0.2 *T/min* (red curve, open circles). Inset (a) shows $\Delta M_{patt}/\Delta M_{unpatt}$ as a function of β at 0.2 *T*. Inset (b) shows R vs. logβ at 0.2 *T* for the patterned sample (for convenience in taking logβ, we used Oe/sec units).

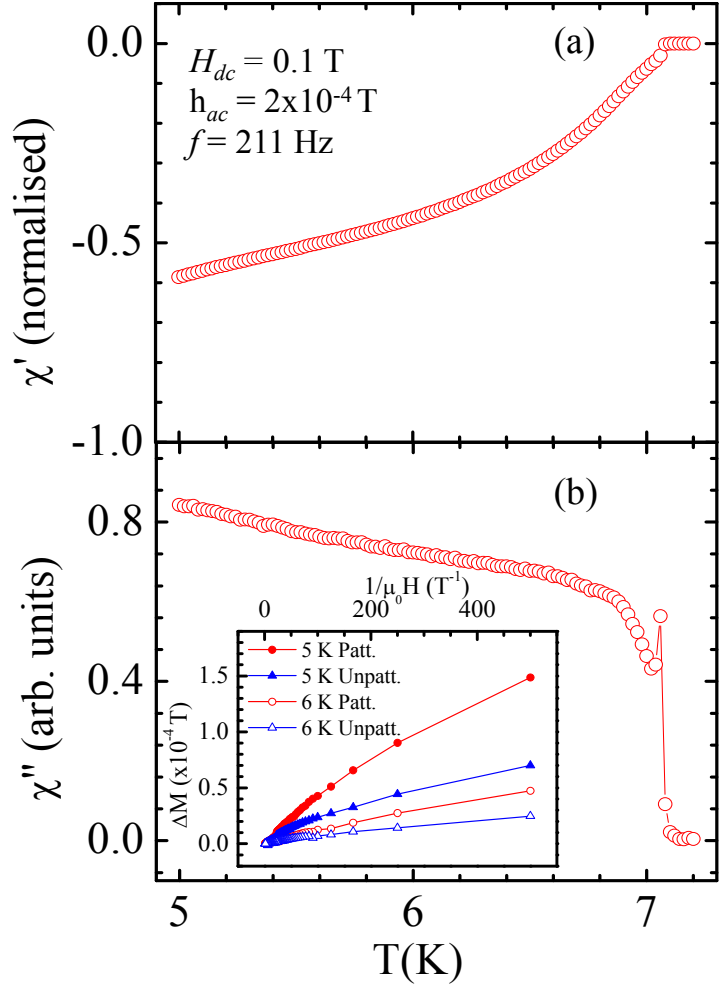

**Fig.3:** (a) and (b) show the in-phase ($\chi'$) and out of phase component ($\chi''$) respectively of the ac – susceptibility response measured as a function of T in a high dc field of $H_{dc} = 0.1$ *T* with a superimposed probing ac field ($h_{ac}$) of $2 \times 10^{-4}$ *T* (frequency 211 *Hz*) for the 2H-NbSe$_2$ crystal prior to patterning. The inset of (b) shows the variation of the width of the M(H) loop ($\Delta M$) versus $\frac{1}{H}$ for the patterned and unpatterned samples at 5 *K* and 6 *K*.

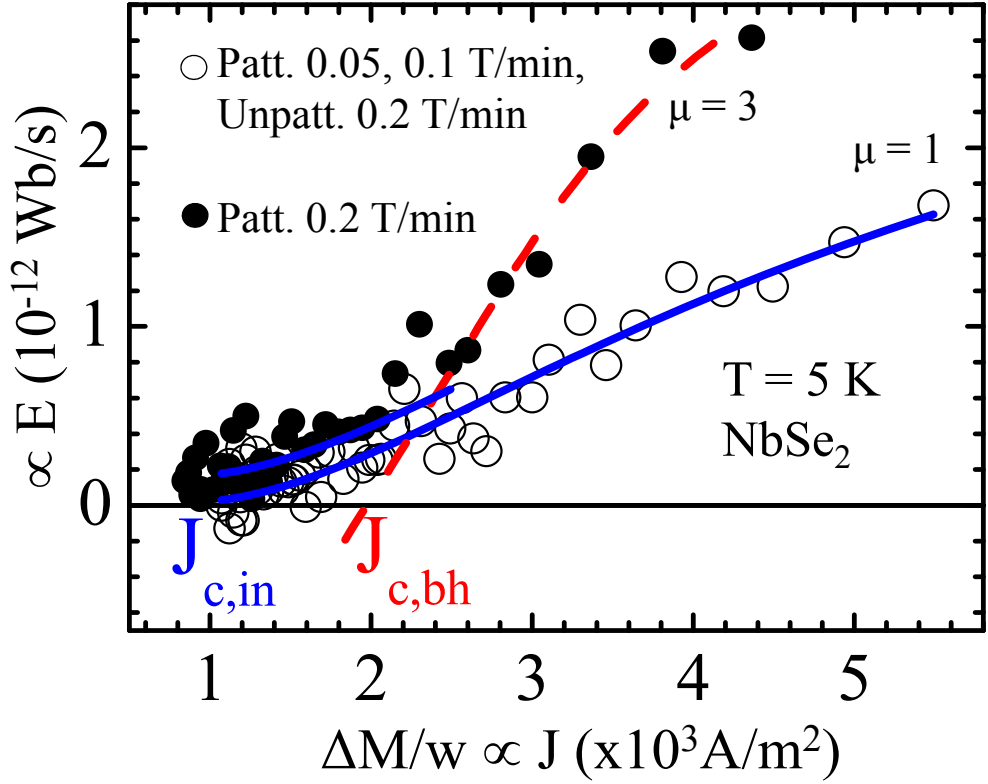

**Fig.4:** [A × (dM$_{for}$/dH)/dH/dt)] (labeled as ∝ E) versus (ΔM/w) for different field sweep rates. Open and closed circles correspond to data for the patterned sample at 0.05 and 0.1 *T/min* and 0.2 *T/min* respectively. The solid (blue) and dashed (red) lines are fit to the data with $\mu = 1$ and $\mu = 3$ respectively. Note that the solid (blue) curve also fits to the data for the unpatterned sample for 0.2 *T/min*.

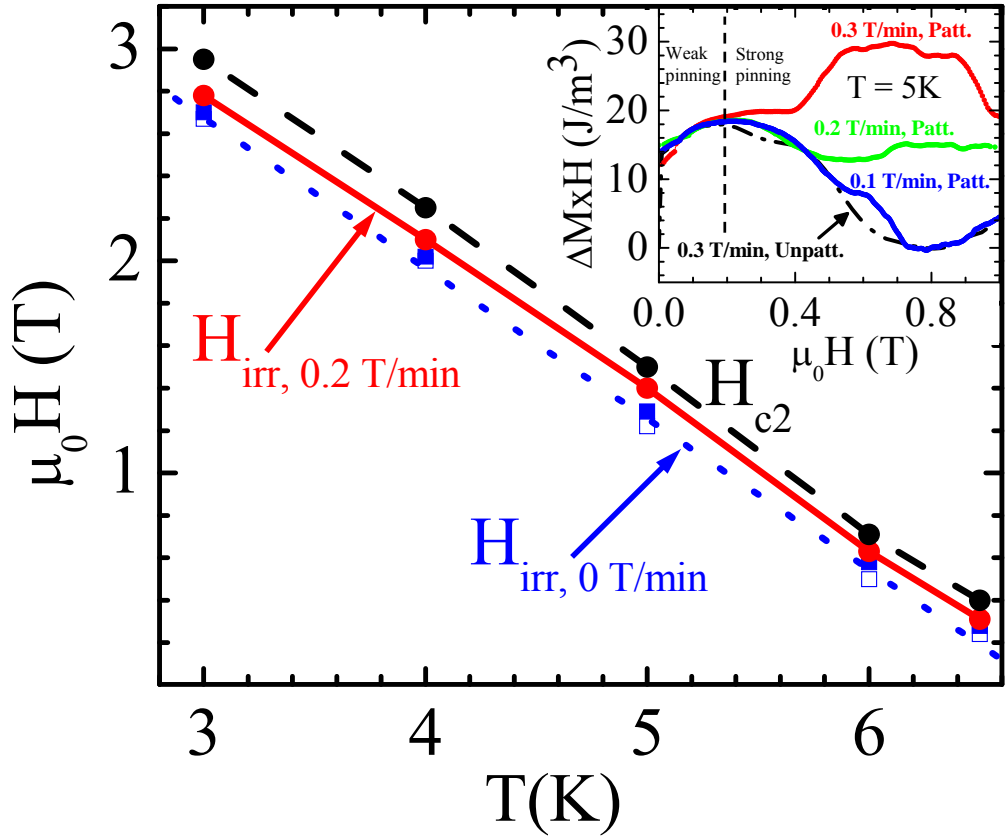

**Fig.5:** Comparison of $H_{irr}(T)$ line measured from hysteresis loops recorded with the field being swept at 0.2 *T/min* (solid (red) line) ($H_{irr,\ 0.2T/min}$) and at 0.005 *T/min* (dotted (blue) line) ($H_{irr,0T/min}$). The inset shows the comparison of the pinning force ($F_P$ *vs* $H$) curve for the unpatterned (recorded at 0.3 *T/min*, black dashed curve) and patterned (at different sweep rates, solid (colored) curves) samples.